\documentclass[11pt]{article}
\pdfoutput=1

\usepackage{amsmath, amsfonts, amssymb}
\usepackage{comment}
\usepackage{graphicx}
\usepackage{psfrag}
\usepackage[usenames,dvipsnames,svgnames,table]{xcolor}
\usepackage{enumerate}
\usepackage{soul}
\usepackage{subfig}
\usepackage{cite}
  \usepackage{a4wide}
  
  \usepackage{color}
  \definecolor{dark-gray}{gray}{0.20}
  \definecolor{gray}{gray}{0.30}
  \definecolor{light-gray}{gray}{0.80}
  \definecolor{dark-red}{rgb}{0.7,0,0}
  \definecolor{dark-green}{rgb}{0.1,0.4,0}
  \definecolor{dark-blue}{rgb}{0.3,0.3,0.7}
  \definecolor{light-blue}{rgb}{0.8,0.8,1}

\usepackage{setspace}

\newcommand{\be}{\begin{equation}}
\newcommand{\ee}{\end{equation}}

\def\be{\begin{equation}}
\def\ee{\end{equation}}
\def\bea{\begin{eqnarray}}
\def\eea{\end{eqnarray}}

\newcommand{\e}{\textrm{e}}

\def\siml eqref{\; \raise0.3ex\hbox{$<$\kern-0.75em
      \raise-1.1ex\hbox{$\sim$}}\; }
   \def\simg eqref{\; \raise0.3ex\hbox{$>$\kern-0.75em
      \raise-1.1ex\hbox{$\sim$}}\; }

\numberwithin{equation}{section}
\usepackage{hyperref}
\hypersetup{
	colorlinks=true,
	linkcolor=dark-blue,
	citecolor=dark-red,
	urlcolor=dark-green,
	linktoc=page
}

\begin{document}

\begin{center}

{\LARGE {\bf A dS obstruction and its \vspace{0.3cm}\\ phenomenological consequences}} \\
\vspace{0.5cm}

\vspace{1.5 cm} {\large  Miguel Montero,$^{a}$  Thomas Van Riet$^{b}$ and Gerben Venken$^{b}$ }\\
\vspace{0.5 cm} 
\vspace{0.5 cm}  \vspace{.15 cm} {$^{a}$Jefferson Physical Laboratory, Harvard University\\
Cambridge, MA 02138, USA}\\ \vspace{.15 cm} 

{$^{b}$Institute of Theoretical Physics, KU Leuven,\\
Celestijnenlaan 200D B-3001 Leuven, Belgium
}

\vspace{0.7cm} {\small \upshape\ttfamily  mmontero @ g.harvard.edu,  thomas.vanriet, gerben.venken @ kuleuven.be
 }  \\

\vspace{2cm}

{\bf Abstract}
\end{center}
{In this note we observe that positive runaway potentials can generically be stabilized by abelian $p$-form fluxes, leading to parametrically controlled  de Sitter solutions after compactification to a lower dimension. When compactifying from 4d to 2d the dS solutions are metastable, whereas all higher dimensional cases are unstable. The existence of these dS solutions require that a certain inequality involving the derivatives of the potential and $p$-form gauge coupling is satisfied. This inequality is not satisfied in simple stringy examples (outside of the scope of Maldacena-Nu{\~n}ez), which unsurprisingly avoid this route to dS solutions. We can apply our techniques to construct $dS_2$ solutions in the Standard Model plus an additional runaway scalar such as quintessence. Demanding that these are avoided leads to (weak) phenomenological constraints on the time variation of the fine structure constant and QCD axion-photon coupling.}

\setcounter{tocdepth}{2}
\newpage

\tableofcontents

\vspace{1cm}

\section{Introduction}

One of the most important open questions in string theory is the quest for a de Sitter solution. While there is a vast literature on precise and concrete strategies to construct a dS solution in string theory, dating back at least to \cite{Kachru:2003aw}, our own point of view is that de Sitter constructions in string theory have never been fully explicit and trustworthy. One could therefore entertain the logical possibility that there is something about string theory that makes it incompatible with meta-stable de Sitter vacua \cite{Danielsson:2018ztv}. This has been turned into a conjecture in \cite{Obied:2018sgi}\footnote{See also \cite{VanRiet:2011yc, Danielsson:2018ztv, Brennan:2017rbf} for earlier conjectures that there is no dS in string theory.} which furthermore constraints the effective potential that any EFT derived from string theory should obey. There is no rigorous argument for the conjecture so far (see however \cite{Ooguri:2018wrx} and \cite{Hebecker:2018vxz} for connections to the Swampland Distance Conjecture \cite{Ooguri:2006in}). As stressed in \cite{Andriot:2019wrs, Junghans:2018gdb} such a general argument might be the reason behind the failure of finding weakly coupled tree-level dS vacua in \cite{Caviezel:2008tf, Danielsson:2011au, Roupec:2018mbn}. 

The inspiration for this conjecture comes from the general scaling properties of \emph{tree-level} flux compactifications derived earlier in \cite{Danielsson:2009ff,Wrase:2010ew,VanRiet:2011yc} which in turn took inspiration from \cite{Silverstein:2007ac, Hertzberg:2007wc}. This does not constitute real evidence since there is no immediate reason why de Sitter vacua should hide in the classical corner of the landscape. However there is no obvious theorem forbidding vacua in that corner either (see \cite{Grimm:2019ixq} for recent progress in this direction). The fact that tree-level solutions, regardless of whether they are free of tachyons, have only been found at strong coupling (see \cite{Caviezel:2008tf, Danielsson:2011au, Roupec:2018mbn} for a scan and overview of such solutions) might be suggestive of a deeper reason. Especially the arguments of \cite{Junghans:2018gdb} seem to hold quite generally and are discouraging to find weakly coupled solutions\footnote{Although there is a recent revival of this discussion since classical vacua could be around at weak coupling once we allow certain singularities \cite{Cordova:2019cvf}, which however have been questioned to be physical \cite{Cribiori:2019clo}.}. 

Of course, from these observations one cannot jump to conclusions about the non-existence of de Sitter vacua in the intermediate regime. At the same time one cannot claim that from a ``potpourri''  of quantum corrections to 10d supergravity one arrives at flux vacua which can be trusted sufficiently to carry the burden of proving a landscape/multiverse charachter of string theory. Even more, there are also non-trivial dS no-go's in string theory that go well beyond 10d tree-level: various papers have studied the effect of an infinite tower of higher derivative corrections, either from the worldsheet or in genuine top down effective field theory, still coming to a no-dS result \cite{Green:2011cn, Gautason:2012tb, Kutasov:2015eba, Quigley:2015jia}. So it seems that in all corners where reliable computations can be done de Sitter escapes existence. 

In any case, a no-dS viewpoint leads one to naturally focus on runaway potentials \cite{Agrawal:2018own} in string theory and to quintessence models in phenomenology \cite{Caldwell:1997ii, Peebles:2002gy}  (see \cite{Copeland:2006wr} for a review). Embedding quintessence in string theory is also difficult and we will not go into any detail of this, but instead refer to \cite{Cicoli:2018kdo, Hebecker:2019csg, Olguin-Tejo:2018pfq}.

The simple point we make on this note is that runaway potentials together with abelian $p-1$-form gauge fields (ubiquitous in string compactifications) naturally lead again to (lower-dimensional) de Sitter solutions. Flux compactifications can stabilize the runaway potential, leading to $dS_{D-p}\times S^p$ solutions. These solutions are under parametric control by taking the flux threading the $S^p$ to be large. Hence they would be an example of explicit, controlled, simple dS saddle points, similar to the usual flux compactifications leading to AdS vacua. We will find that for $D-p>2$ the solutions are generically unstable, but for $D-p=2$, the instability in the homogeneous mode disappears.  In fact, in the two-dimensional case, this solution arises as a near-horizon limit of charged solutions in the higher-dimensional theory, just like their AdS counterparts. In fact this work first arose as an attempt to extend charged Nariai black hole solutions which are $dS_2\times S^2$ describing the near-horizon geometry of large black holes in de Sitter space, to slow-roll quintessence spacetimes.

In this paper we just work out the details, and find that, in order for a $dS_{D-p}\times S^p$ saddle point to exist, it is necessary and sufficient that a simple certain inequality involving the gauge kinetic function of the $p-1$-form field and the potential is satisfied. More concretely, if $f$ is the inverse gauge coupling of the $(p-1)$-form field, dS saddle points exist provided that
\begin{equation}(p-1)\left\vert\frac{V'}{V}\right\vert \leq\left\vert\frac{f'}{f}\right\vert \label{dshigh0}.\end{equation}

We then check against simple controlled examples, including the $O(16)\times O(16)$ nonsupersymmetric string, and find that the inequality is not satisfied. If someone ever comes up with a worldsheet de Sitter vacuum, our techniques could have been used in principle be used to stabilize the dilaton, thereby turning these constructions into actual quantum gravitational dS saddle points\footnote{Again, by ``dS saddle points'' we mean that our higher dimensional $dS$ solutions are unstable, while the 2-dimensional ones are not.}. We check explicitly that this doesn't work, either. The inequality we have is marginally violated,\ the amount of the violation is parametrically small and goes as $g_s$. 

As emphasized above, for the 2-dimensional case we get stable $dS_2$ solutions\footnote{As we discuss below, there is a ``Bousso beading process'' inhomogeneous instability, which we interpret as nucleation of two-dimensional black holes by the expanding $dS_2$ background. This instability can never destroy the $dS_2$ completely (at any given time, we expect most of the spatial volume is in a $dS_2$ patch).}. The instabilities are frozen via the Hamiltonian constraint. If the no dS or any similar conjecture on the string potential (such as TCC \cite{Bedroya:2019snp}) are true, then the opposite of \eqref{dshigh0} is a logical consequence. Absence of dS solutions implies constraints on the couplings in a different, runaway solution. 

It turns out that simple extensions of the Standard Model with scalars, such as for instance quintessence or a QCD axion, fall into our framework, and therefore the no dS conjecture constrains\footnote{We would constrain by demanding absence of $dS_2\times S^2$ solutions, which is fishy due to the peculiarities of low-dimensional quantum gravity. For instance, \cite{Maldacena:2019cbz} proposes a version of the SYK model which is dual to a patch of a (non-Einstein) $dS_2$ quantum gravity. On the other hand, the kind of arguments ruling out tree-level higher-dimensional dS compactifications in string theory seem to apply equally well to 2d.}  these sectors via (the opposite of) \eqref{dshigh0}. In particular, the time-variation of the fine structure constant is upper-bounded by the quintessence slow-roll parameter $\vert V'/V\vert$. Current experimental upper bounds on this, coming from the Oklo natural nuclear reactor \cite{Damour:1996zw}, ensure that this constraint is satisfied in the real world, for $\vert V'/V\vert$ as small as $10^{-5}$. We find it interesting that the no-dS conjecture, favoring quintessence, is only consistent with a weak time-variation of the fine-structure constant, which otherwise plagues quintessence models. 

Another interesting consequence comes from analyzing the photon-axion coupling, which is the subject of active experimental searches \cite{Tanabashi:2018oca}. In this case, avoiding lower-dimensional de Sitter solutions requires the product of the axion-photon coupling and the axion decay constant to be a not-too-large number. While this constraint is of no practical phenomenological interest, we find it interesting that it exists. It may point out the way to further, more interesting constraints.  

The note is organized as follows. In Section \ref{Sec:Nariai} we recall charged Nariai black holes in Einstein-Maxwell theory since their near horizon is $dS_2\times S^2$, and serve as a prototype for the kind of solutions we are interested on. In Section \ref{Sec:4d2d} we present our construction in any dimension, illustrating the details in the four-to-two dimensional case, write down the constraints that must be satisfied for the solution to exist, and compare to stringy wisdom. The phenomenological implications of our results + the no-dS conjecture are then analysed in Section \ref{sec:pheno}. We conclude and discuss our assumptions in Section \ref{Sec:discussion}, leaving the rest of the technical details to Appendices.

\section{Warm up: charged black holes in dS space}\label{Sec:Nariai}
We will start by describing the charged black hole solutions of the Einstein-Maxwell-theory, with a particular focus on Nariai black hole solutions. The low-energy Lagrangian we will take is
\begin{equation}
S=\int d^4x\sqrt{-g}\left[ \frac{1}{16\pi G}\left(\mathcal{R} -\frac{6}{\ell^2}\right) - \frac{1}{4g^2}F_{\mu\nu}F^{\mu\nu} \right].\end{equation}
This theory allows a  range of spherical black hole solutions called charged Reissner Nordstr\"om \cite{Romans:1991nq} and are of the form
\begin{equation}
ds^2=-U(r) dt^2+\frac{dr^2}{U(r)} +r^2 d\Omega,\label{RNdSmetric}
\end{equation}
with
\begin{equation}U(r)\equiv 1-\frac{2GM_r}{r} + \frac{G(gQ_r)^2}{4\pi r^2}-\frac{r^2}{\ell^2}.\label{Ur0}\end{equation} 
This metric can either describe magnetic or electric charge.  For the electrically charged solution we have 
\begin{equation}
A=\Phi\, dt, \quad \Phi=\frac{g^2}{4\pi} \frac{Q_r}{r}.\label{A}
\end{equation}
The two parameters $M_r$ and $Q_r$ can be interpreted as a ``mass'' and the charge. But in de Sitter space the notion of mass is ambiguous.

\begin{figure}[!htb]
	\begin{center}
		\includegraphics[width=0.55\textwidth]{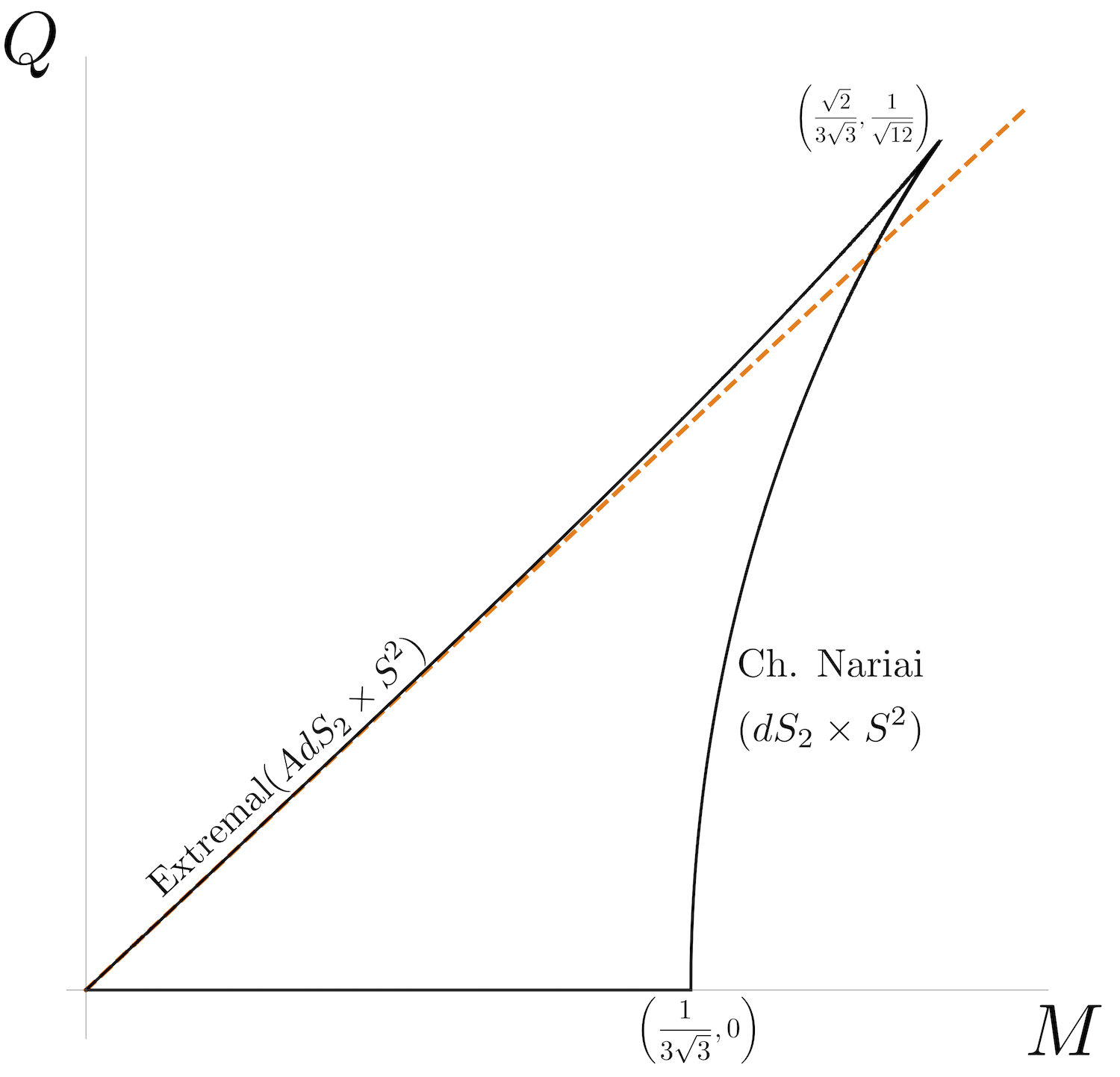}
		\vspace{0.3cm}
		\caption{\small{Phase diagram of Reissner-Nordstrom-de Sitter black holes with the dimensionless parameters $M\equiv \frac{GM_r}{\ell},\quad Q^2\equiv \frac{Gg^2Q_r^2}{4\pi \ell^2}$. Sub-extremal solutions only exist inside the shark-fin shaped region. The boundary of this allowed region has two branches: On the top branch one has extremal RN-dS black holes, which have a $AdS_2\times S^2$ horizon topology. On the lower right branch one has charged Nariai black holes, charged dS black holes in which the black hole and cosmological horizons coincide. These have a $dS_2\times S^2$ near-horizon geometry. The orange dashed line is the ``lukewarm line'' $Q=M$, where the surface gravity (and hence temperature) of the two horizons are identical.}}\label{fig1}
	\end{center}
\end{figure}

The RN-dS metric features the event horizon of the charged black hole but there is also the cosmological horizon which causes black holes in dS space to have a maximal size.  Figure \ref{fig1}, taken from \cite{Montero:2019ekk}, shows the space of solutions.  The boundary of the diagram comprises two different branches \cite{Romans:1991nq}:
\begin{itemize}
	\item The upper branch of the black curve in Figure \ref{fig1} parametrizes extremal black holes at zero temperature with horizons smaller than the cosmic horizon. The inner and outer black hole horizon coincide, $r_+=r_-$. The near-horizon geometry is $AdS_2\times S^2$. 
	\item The lower branch of the black curve in Figure \ref{fig1} is the charged Nariai branch, and it contains subextremal charged black holes with the same area as the cosmological horizon.  The near-horizon geometry is $dS_2\times S^2$.
\end{itemize}
These two branches meet at a point with vanishingly small temperature where the near horizon geometry is $\mathbb{M}_2\times S^2$. Outside of the diagram one always has naked singularities, except on the $Q=0$ axis of neutral black holes, where one has a Big Crunch singularity instead. 
 
The boundary of the diagram is therefore given by the set of ``vacuum solutions" in the 2d theory obtained from reducing over the $S^2$. In the next section we investigate such solutions for Einstein-Maxwell-quintessence theories.

\section{Constructing de Sitter from runaway potentials}\label{Sec:4d2d}

One can wonder whether a phase diagram similar to charged black holes in dS$_4$ exists for black hole solutions in quintessence spacetimes. This is much more involved questions since we are not aware of exact solutions for black holes. But for the purpose of this note we care about the boundary of the would-be phase diagram only, and more concretely the $dS_2$ branch. Hence our approach consists of only looking for flux compactification solutions of the form $M_2 \times S^2$ with $M_2$ being maximally symmetric. For related work see \cite{Bousso:1996pn}.

From here onward we work in units where $16\pi G=1$ and consider the following action for the quintessence and electromagnetism coupled to gravity
\be \label{action}
S= \int\sqrt{|g|}\Bigl(\mathcal{R}-\tfrac{1}{2}(\partial\phi)^2 -\tfrac{1}{4}f(\phi) F_{\mu\nu}F^{\mu\nu} - V(\phi)\Bigr)+\frac{\theta}{8\pi^2}  F\wedge F \,.
\ee
Here $V$ is the quintessence potential and $f$ the coupling to electromagnetism. We have also allowed for a $\theta$ angle which can depend on $\phi$; this allows us to cover the case $\theta=\frac{n}{f_\phi}\phi$ when the quintessence field $\phi$ is itself an axion, but for the time being we keep the coupling general.

We search for solutions with a product spacetime $M_2 \times S^2$ endowed with the metric
\be\label{Ansatz}
d s_4^2 = h(t)[-d t^2 + dr^2] + R^2d\Omega_2^2\,.
\ee
We now check whether we can recover the horizon geometries that exist without quintessence. For that consider the scalar equation of motion:
\begin{equation}
\frac{\partial_{\mu}\bigl( \sqrt{|g|}\partial^{\mu} \phi\bigr)}{\sqrt{|g|}} = V' + \tfrac{1}{4}f'F_{\mu\nu}F^{\mu\nu}-\frac{\theta'}{8\pi^2} \frac{1}{4\sqrt{|g|}}\varepsilon^{\mu\nu\rho\sigma}F_{\mu\nu}F_{\rho\sigma}\,,
\label{eqphi}\end{equation}
where a prime denotes a derivative with regards to to the scalar and $\varepsilon$ is the anti-symmetric symbol (not tensor).  We restrict to solutions with $\partial^\mu \phi = 0$. We will turn to the stability of such homogeneous solutions against perturbations in section \ref{sec:stabilityanalysis}\,. The scalar equation of motion then becomes
\begin{equation}
0 = V' + \tfrac{1}{4}f'F_{\mu\nu}F^{\mu\nu}-\frac{\theta'}{8\pi^2} \frac{1}{4\sqrt{|g|}}\varepsilon^{\mu\nu\rho\sigma}F_{\mu\nu}F_{\rho\sigma}\,,
\label{stabilizedscalarthingum}\end{equation}

Let us next address Maxwell's equations
\begin{equation}
d\left(*fF-\frac{1}{4\pi^2}\theta F\right)=\delta_E ,\quad d F=\delta_M.\end{equation}
Writing $F= E(r)dr\wedge dt+B(r)R^2d\Omega$, one gets from the above, assuming the scalar has been stabilised \eqref{stabilizedscalarthingum},  
\begin{equation} E=\left(Q_e+\frac{\theta}{4\pi^2} Q_m\right)\frac{h}{fR^2},\quad B= \frac{Q_m}{R^2}\,.\end{equation}
Where charges were defined, via conserved currents, using the following normalisation:
\be
Q_e = \frac{1}{4\pi}\int_{S^2} \left( f\star F - \frac{\theta}{4\pi^2} F\right)\qquad  Q_m = \frac{1}{4\pi}\int_{S^2} F\,.
\ee
The charge $Q_e$ is integer quantized and $Q_m$ is quantized in multiples of $2\pi$.
For convenience we introduce the following two quantities:
\begin{align}
& \tilde{Q}_e=Q_e + \frac{\theta}{4\pi^2}Q_m\\
& \mathcal{Q}^2 = \frac{\tilde{Q}_e^2}{f} + fQ_m^2\,,
\end{align}
The equations of motion for a constant scalar $\phi$ and radius $R$ take then a simple form 
\begin{align}
& V' + \frac{1}{2R^4}{\mathcal{Q}^2}' = 0\,,\label{eq1}\,\\
& 0 = VR^4 -2R^2 +\frac{1}{2}\mathcal{Q}^2\label{eq2}\,,\\
& \mathcal{R}_2 = V - \frac{1}{2R^4}\mathcal{Q}^2 \label{eq3}\,.
\end{align}
where $\mathcal{R}_2$ is the 2d curvature\footnote{$\mathcal{R}_2 =-h^{-2}\ddot{h}+h^{-3}\dot{h}^2$.}. A derivation of these equations can be found in Appendix \ref{App:EOM}.

Equation (\ref{eq1}) stabilises the scalar field for a given $Q_e$ and $Q_m$ and $V$. If that equation has a solution then the functions $f$ and $V$ obtain specific fixed values. One would then be tempted to think that the remaining two equations are then similar to the Einstein-Maxwell-deSitter case upon the identification: $V= 6\ell^2$ and $f=g^2$ (recall we work in Planckian units for which $16\pi G=1$). But this turns out not to be the case, probably due to the fact that the the stabilised scalar value still depends on $R$.

In any case, the existence of solutions depends on the value of $\mathcal{Q}^2$ and $V$ at the stabilised value of the scalar (if any). From equation (\ref{eq2}) it is clear that a real solution for $R$ implies that
\be\label{condition0}
V\mathcal{Q}^2 \leq 2\,.
\ee

Using \eqref{eq1}, \eqref{eq3} becomes
\begin{equation}\mathcal{R}_2=V\left(1+\frac{V'}{V}\frac{\mathcal{Q}^2}{(\mathcal{Q}^2)'}\right)\,.\end{equation}
If the potential $V$ is positive, then using \eqref{eq1} again to show that $V'/V$ and $(\mathcal{Q}^2)'/\mathcal{Q}^2$ have opposite signs, we will have a $dS_2$ solution only if
\begin{equation}\left\vert\frac{V'}{V}\right\vert\leq \left\vert \frac{(\mathcal{Q}^2)'}{\mathcal{Q}^2}\right\vert.\end{equation}
The right hand side becomes a sum of two terms
\begin{equation}\frac{(\mathcal{Q}^2)'}{\mathcal{Q}^2}=\frac{f'}{f}\frac{f^2Q_m^2- \tilde{Q}_e^2}{\tilde{Q}_e^2+f^2Q_m^2}+ \frac{2f \tilde{Q}_e Q_m}{\tilde{Q}_e^2+f^2Q_m^2}\frac{\theta'}{4\pi^2 f}.\end{equation}
Since the dependence on the charges is homogeneous, the function above only depends on one parameter, and by explicit computation one can show that \begin{equation}\left\vert \frac{(\mathcal{Q}^2)'}{\mathcal{Q}^2}\right\vert\leq \sqrt{\left(\frac{f'}{f}\right)^2+ \left(\frac{\theta'}{4\pi^2 f}\right)^2},\end{equation}
with the inequality always being achieved\footnote{In the large charge limit, that is, ignoring \eqref{condition0}.} for some value of $\tilde{Q}_e, Q_m$. So without loss of generality, to have a $dS_2$ solution in this way we need to satisfy the constraint
\begin{equation}\left\vert\frac{V'}{V}\right\vert \leq \sqrt{\left(\frac{f'}{f}\right)^2+ \left(\frac{\theta'}{4\pi^2 f}\right)^2}.\label{dScond}\end{equation}

We can immediately generalize the picture to higher dimensions. One can consider a general $D$-dimensional solution with an abelian $p$-form fieldstrength and a scalar with an action of the form
\begin{equation}S=\frac{1}{2\kappa_{D}^2}\int d^{D}x\, \sqrt{\vert g_E^{D}\vert}\left[ R_E-\frac12(\partial\phi)^2-f(\phi)\vert F_p \vert^2-V(\phi)\right].\label{laghd}\end{equation}
Consistent truncations of this kind often show up in string compactifications after supersymmetry is broken and some of the moduli develop a potential. In this case $f$ and $V$ are often exponentials. The detailed analysis, which can be found in Appendix \ref{app:higherdim}, leads to the conclusion that there will be a $(D-p)$-dimensional $dS$ solution obtained from flux compactification if and only if

\begin{equation}(p-1)\left\vert\frac{V'}{V}\right\vert \leq \left\vert\frac{f'}{f}\right\vert \label{dshigh}.\end{equation}

\subsection{Classical limit }
We need to check that the $dS$ solutions obtained in this way are in the semiclassical approximation. In the two-dimensional case, this means that $R$ has to be large, $\mathcal{R}_2$ small, and the charges large enough such that quantization is not a worry. The detailed analysis can be found in Appendices \ref{App:EOM} and \ref{app:higherdim}, with the conclusion being that there is always a large charge regime where the solution is trustworthy.

Quantum corrections will appear in general Casimir energy contributions to the lower-dimensional potential. Again, the Nariai solution gives an intuition as to why these are small: the $S^2$ of a Nariai black hole is as big as the parent universe, and such backgrounds are essentially classical if the parent de Sitter solution is classical as well. More concretely, a Casimir energy in lower dimensions will go as $H^{D-p}$, to be compared with the classical contribution to the potential, which is of order $V_0/H^p$. Here we are assuming the energy density is of order $V_0$, since we want to stabilize the scalar at the same value that it has in the parent solution, and we are integrating over the $S^p$. But $H^2\sim V_0$ due to the higher dimensional Einstein's equations, and so
\begin{equation} V_{\text{Casimir}}\sim H^{D-p}\ll H^{2-p}\,,\end{equation}
for small $H$ as long as $D>2$. 

We also point out that, in contrast to the two-dimensional case, for higher dimensions our solutions are controlled against any higher-derivative corrections since we can make the charge parameter arbitrarily large, so that the solutions can be arbitrarily weakly curved.

\subsection{Stability}\label{sec:stabilityanalysis}
Finally we address concerns about stability. Clearly stability depends on the particular form of $f$ and $V$. When $f$ and $V$ are exponentials, with $f=f_0e^{\delta \phi}$ and $V=V_0e^{-\gamma\phi}$, the solution is always a minimum in the two-dimensional case, but tachyonic in the higher-dimensional case. We expect tachyonic instabilities to be rather generic in the higher-dimensional case. Hence one can contemplate to what extend the dS Swampland conjectures forbid such backgrounds. In here we take a viewpoint that is stronger than what is expressed in \cite{Garg:2018reu,Ooguri:2018wrx,Ooguri:2018wrx}; as long as there are no explicit examples of unstable dS solutions at the classical level that are under parametric control there might be a good reason to suspect that such backgrounds are equally forbidden. Note that all examples constructed so far of classical tachyonic dS solutions from fluxes and orientifolds \cite{Danielsson:2011au,Roupec:2018mbn, Junghans:2018gdb} are either at strong string coupling or at strong curvature, meaning they are not even approximate solutions to string theory. If string theory describes the real world then one could consider the maximum of the Higgs potential as a counterexample \cite{Denef:2018etk}, but perhaps this solution does not live at parametric weak coupling.

Wether a $dS_2$ solution obtained from a flux compactification is unstable is a subtle question. As we discuss in Appendix \ref{App:EOM}, stability of 4d scalars, such as $\phi$, is studied via positivity of the Hessian as usual. However, stability of the scalar $R^2$, which parametrizes the size of the $S^2$, is more subtle.  As explained in \cite{ArkaniHamed:2007gg}, and analyzed in detail in \cite{Hamada:2017yji}, there are no instabilities in the homogeneous mode, where $R^2$ is perturbed in a position-dependent way on a a $t=0$ slice of global $dS_2$ coordinates. This is because the Hamiltonian constraint of GR fixes $R^2$ as a function of the other parameters of the theory, so it cannot fluctuate on its own\footnote{This can be happening at a maximum of the potential, which is  why \cite{ArkaniHamed:2007gg} talks about ``top-of-the hill'' but stable $dS_2$ models.}.

However, as first explained in \cite{Bousso:1998bn} (see also \cite{Niemeyer:2000nq,Bousso:2002fq} and especially \cite{Bousso:1999ms}, which discusses the charged case), at the classical level the Hamiltonian constraint allows for inhomogeneous perturbations which are unstable. These perturbations then grow and settle in a different configuration where patches of the $dS_2$ crunch, while others keep on expanding. Thus, the spacetime forms a series of ``beads'', which is why this is called the ``Bousso beading process''. In the charged case, the expanding patches between beads go back to the original $dS_2$ solution. The conformal diagram of the whole situation can be seen in \cite{Bousso:1999ms} and looks like a chain of two-dimensional black holes (the crunching regions) with $dS_2$ patches in between (when the charge is large enough, which is the case we are interested in). The process repeats endlessly, with a proliferation of crunching regions among an ever increasing number of disconnected $dS_2$ patches. 

The Bousso beading process is a universal instability of $dS_2$ models of quantum gravity coming from higher-dimensional compactifications. The question is whether it ``counts'' as an instability for Swampland purposes. That is, would we expect to have  $dS_2$ solutions in string theory, since this instability will be generically present? Most of the no-go results we are familiar with are concerned only with stability in the homogeneous mode, and so would apply in this setup as well. But more importantly, in the beading instability, there are always ever-expanding patches of $dS_2$, so the original vacuum still survives (in the same sense that in eternal inflation almost every patch is in the false vacuum). Finally, to the extent that we refer to the crunching regions as black holes, a natural interpretation of Bousso beading is that it is describing pair production of 2d black holes in the background $dS_2$ vacuum. An expanding universe pair-produces particles of any mass due to the expansion of the universe; we should expect it to pair-produce black holes as well. If this interpretation is correct, Bousso beading is not a vacuum instability any more than particle production in any de Sitter background (also higher-dimensional ones) is.

\subsection{Checking against string theory}\label{sec:strr}

The main point of this note, explained above, is that one can generically get de Sitter solutions under parametric control in a flux compactification of an effective model with a runaway potential. These solutions are akin to the flux compactifications that give rise to $AdS$ solutions in string compactifications, and would similarly correspond to the near-horizon limit of extended charged objects in the higher-dimensional solution with the runaway potential. In one particular case, where $p=D-2$, the solutions we are constructing are actually Nariai-like limits of charged black hole solutions, so this picture is realized exactly. 

For this construction to work, however, \eqref{dScond} or \eqref{dshigh} must be satisfied. So the question is whether one can do this in controlled string compactifications. 

Since the solutions we construct are classical flux compactifications, \eqref{dScond}, or the assumptions leading to it, must be violated whenever the Maldacena-Nu{\~n}ez theorem \cite{Maldacena:2000mw} (and its extensions, see e.g. \cite{Das:2019vnx}) hold, if the single-scalar truncation is consistent. For instance, massive IIA supergravity seems to have all ingredients already in 10D; a runaway exponential potential for the dilaton and a dilaton coupling to the $U(1)$-field:
\be
S_{10}\supset  \int d^{10}x \sqrt{|g_{10}|}\Bigl(R-\tfrac{1}{2}(\partial\phi)^2 -\tfrac{1}{4} e^{3\phi/2}F_2^2 -m^2{\e^{5\phi/2}} \Bigr)\,. \label{eq0}
\ee
We could look for magnetic solutions of the form $dS_2\times S^8$, with $Q$ units of $F_8$ flux threading the $S^8$. However, it turns out that \eqref{dshigh} is not satisfied, since $7\cdot 5/2$ is not smaller than $3/2$. On top of this,  the equation of motion for $H_3$
\begin{equation} d(e^{-\phi} *H_3)= m F_8\label{dl}\,,\end{equation}
means we could not truncate $H_3$. 

As explained above, when we try to run our argument in higher-dimensional field strenghts we actually find perturbatively unstable de Sitter solutions. These are not too interesting anyway (in spite of the observation in the literature that mildly tachyonic dS solutions are not found in the perturbative corners of string theory, or TCC \cite{Andriot:2018wzk,Ooguri:2018wrx,Andriot:2018mav,Bedroya:2019snp}), but it is fun to observe that \eqref{dshigh} is not satisfied in this case either. As a simple example, consider the consistent truncation of the non-SUSY $O(16)\times O(16)$ heterotic string \cite{AlvarezGaume:1986jb} with the $B_2$ field as abelian $p-1$-form fieldstrength. The 10d action in the Einstein frame reads (we have normalized the dilaton so that the string coupling is $e^{-\phi}$)
\begin{equation}S_{E}=\frac{1}{2\kappa_{10}^2}\int d^{10}x\, \sqrt{\vert g_E^{10}\vert}\left[ R_E-\frac12(\partial\phi)^2-e^{\phi}\vert H_3 \vert^2-V_0e^{-\frac{5}{2}\phi}\right],\label{weqe}\end{equation}
where \cite{AlvarezGaume:1986jb}
\begin{equation}V_0=0.03 (\alpha')^5.\end{equation}
The right hand side of (\ref{dshigh}) is 1, and the left hand side is $5$, so no de Sitter solutions are possible. The cosmological constant in ten dimensions only arises as a one-loop effect, which means that the usual $e^{-2\phi}$ factor in the string frame gets cancelled out, and we get the above result. Notice that the fact that there is no dS solution in this case lies out of the scope of the usual Maldacena-Nu{\~n}ez result. While the non-supersymmetric $O(16)\times O(16)$ string can be connected to a non-supersymmetric compactification of M theory in nine dimensions, Maldacena-Nu{\~n}ez refers to classical M-theory compactifications. 

One generically expects runaway potentials whenever supersymmetry is broken, even in situations where Maldacena-Nu{\~n}ez does not apply. Our point is that in such situation, whenever abelian $p$-form fields are present too, there is an easy way to potentially obtain de Sitter solutions. It would be interesting to check whether our constraint applies in more general setups, such as e.g. those discussed in \cite{Grimm:2019ixq}.

We can elaborate a little bit on this case. The $O(16)\times O(16)$ example above failed to produce a (n unstable) de Sitter, but it came quite close. The vacuum energy was generated at one loop, but suppose one was somehow able to produce a model with a tree level contribution to the vacuum energy as well. The potential in \eqref{weqe} then gets replaced as
\begin{equation} V_0 e^{-\frac{5}{2}\phi}\rightarrow V_0 e^{-\frac{5}{2}\phi}+V_{\text{tree}}e^{-\frac{\phi}{2}}.
\end{equation}
Now, for large charge, 
\begin{equation} \left\vert\frac{V'}{V}\right\vert=\frac{(5/2)V_0 e^{-\frac{5}{2}\phi}+(1/2)V_{\text{tree}}e^{-\frac{\phi}{2}}}{ V_0 e^{-\frac{5}{2}\phi}+V_{\text{tree}}e^{-\frac{\phi}{2}}}> \frac12,\label{ssq}\end{equation}
so the inequality \eqref{dshigh} is not satisfied. In fact, the violation is proportional to $e^{-2\phi}$, so it can be made arbitrarily small, but it is always positive.

Nonsupersymmetric flux compactifications of the $SO(16)\times SO(16)$ string have been considered before, in \cite{Basile:2018irz,Antonelli:2019nar}. These references constructed $AdS$ solutions which exist precisely because the inequality \eqref{dshigh} is not satisfied. These references carried out an in-depth perturbation analysis, showing that the $AdS$ solutions are indeed stable. 

A tree-level (worldsheet) de Sitter in string theory\footnote{Even this already faces serious obstacles in heterotic string theory \cite{Kutasov:2015eba}.} would still not be a de Sitter compactification of quantum gravity, since at any finite value of the coupling there would be a runaway potential for the dilaton. Even so, one could have tried to stabilize it with our techniques to turn this tree-level de Sitter into a different, exact dS solution, via a flux compactification. This would be an example of a parametrically controlled dS saddle point in string theory. Equation \eqref{ssq} shows that even this cannot happen.

It therefore seems that our statement \eqref{dshigh} is verified in string theory, even though when violations would only lead to a tachyonic dS, which doesn't seem too bad. One possibility is that even mildly tachyonic dS solutions are not allowed in string theory \cite{Andriot:2018wzk,Ooguri:2018wrx,Andriot:2018mav}. Another possible explanation is that it might be possible to further compactify the higher-dimensional solutions to produce $dS_2$ solutions, where the instability is absent \cite{Brown:2013mwa}. We leave exploration of this to future work.

We also need to address the Swampland Distance Conjecture \cite{Ooguri:2006in}, since generically our $dS$ solutions sit at a certain value of $\phi$ that depends on $Q$. For us to be able to trust the analysis for abitrarily large $Q$, the Lagrangian descriptions \eqref{action} or \eqref{laghd} should be valid in the deep IR for large values of $\phi$. This is the case in some examples, like the $O(16)\times O(16)$ perturbative heterotic string discussed above, where the only light fields are the dilaton, metric, NSNS field, gauge fields (which can be consistently truncated to zero) and fermions. Everything else is at least string-scale heavy. While the tower of states coming from stringy modes comes down in Planck units as $\phi\rightarrow\infty$, the solutions we look at are always within the validity of the EFT. 

\subsection{The web of conjectures is self-consistent}\label{Sec:web}
There is one last thing we need to check. Our claim is that \eqref{dScond} holds for scalars with an arbitrary potential $V(\phi)$. So what about any of the very massive string or KK modes that are ubiquitous in string compactifications? When $V=1/2 m^2\phi^2$, we will get a dS solution only if
\begin{equation} \phi\geq\frac{2(p-1)}{\vert f'/f\vert }.\end{equation}
That is, the $dS$ solution is at a displaced value of the scalar. If this is a stringy or KK mode, we expect coupling to additional modes to spoil the analysis anyway, but it is interesting that for $f'/f$ of order one the value of $\phi$ is Planckian. This is a significant distance away and it is very unclear one should trust the effective field theory in this case.

In the above simple analysis we assumed that the vacuum energy at $\phi=0$ vanishes. It is also interesting to see what happens in the $AdS$ case.  For that consider an action of the form (\ref{action}) with a potential that has a non-tachyonic $AdS_4$ minimum:
\be
V(\phi) = -\frac{1}{L_{AdS}^2} + m^2\phi^2 + \mathcal{O}(\phi^3)\,.
\ee 
Now if $f'/f$ is order one we  can try to construct a $dS_2$ by stabilizing $\phi$ away from its $AdS_4$ minimum using electromagnetic fluxes in such a way that the 4d potential in that part of moduli space is positive. Then we will have a $dS_2$. Interestingly, for that to happen the scalar needs to travel at least
\begin{equation}
\Delta \phi > \frac{1}{Lm}\,. 
\end{equation}
The distance conjecture \cite{Ooguri:2006in} however tells us that this should not be super-Planckian since otherwise the EFT is not to be trusted \cite{Blumenhagen:2018hsh}. So we have that we get a de Sitter vacuum in 2d if
\begin{equation}\label{AdScale}
Lm>1\,,
\end{equation}
but this violates the AdS-moduli scale separation conjecture \cite{Gautason:2018gln}, which is a weaker form of the AdS-KK scale separation conjecture of \cite{Lust:2019zwm} (see also \cite{Gautason:2015tig}).  Note that satisfying (\ref{AdScale}) was conjectured to be impossible exactly because it would lead to dS vacua in 4D after SUSY breaking, quasi-independent of the details of that breaking \cite{Gautason:2018gln}. Here we arrive at the same condition from demanding the absence of de Sitter vacua after an $S^2$ compactification.  

So the assumption of both the distance conjecture and the more speculative no-dS conjecture implies the AdS scale separation condition. The only thing we assumed is that $f'/f$ is order 1, which seems true for the dilaton or volume modulus in string theory. 

\section{Phenomenological implications}\label{sec:pheno}
In the previous section, we have explained how one can generically construct de Sitter solutions from flux compactifications where there is a scalar with a runaway potential, and how this technique resists application to string theory. Our analysis has been completely rigorous up to this point, but we now switch to conjecture mode. Reference \cite{Obied:2018sgi} (see also \cite{Danielsson:2018ztv}) raised the possibility of no de Sitter solutions in string theory. Coupled to our observations above, this can only be possible if in any string compactification, the converse of \eqref{dshigh} is satisfied. In other words, whenever a string compactification leads to a low-energy effective field theory such that \eqref{laghd} is a consistent truncation, one must have

\begin{equation}(p-1)\left\vert\frac{V'}{V}\right\vert \geq \left\vert\frac{f'}{f}\right\vert \label{dshigh2}.\end{equation}

So this is a concrete prediction of the no dS conjecture for string compatifications, but there is more to say. If the real world is indeed described by string theory, and the conjecture is correct, it follows that the presently observed positive dark energy must be in some form of quintessence. But then, our construction in the previous Section works and gives us a two-dimensional $dS$ solution.  As explained above, this should be thought of as the near-horizon limit of a charged Nariai black hole. There are two possibilities:
\begin{itemize}
\item The no $dS$ conjecture or its variants do not apply in two dimensions. Gravity in two-dimensions is qualitatively different and some Swampland constraints can be absent (e.g. there are global symmetries in the string worldsheet).
\item The conjecture should also apply. After all, the kind of arguments we have nowadays to rule out dS solutions in corners of moduli space come essentially from dimensional reduction of 10d string theory, and do not treat the two-dimensional case differently in any obvious way.
\end{itemize} 
Modulo subtleties about stability in 2D and the Bousso beading process, we will now explore the consequences of the second possibility. We will go back to the scenario described in Section \ref{Sec:4d2d}, but take the gauge field to be the electromagnetic $U(1)$ we see in the real world. $\phi$ will be the quintessence field. Some references which look at two-dimensional compactifications of the Standard Model are \cite{ArkaniHamed:2007gg,Arnold:2010qz,Ibanez:2017oqr,Hamada:2017yji}. We just add a slowly rolling scalar to the mix. In particular, \cite{Arnold:2010qz} describes the cousin $AdS_2\times S^2$ compactifications describing the near-horizon limit of an extremal but large RN black hole in $dS_4$. To avoid a $dS_2$ solution, we must violate either \eqref{condition0} or \eqref{dScond}. Equation \eqref{condition0} depends only on the current-day values of the vacuum energy and $f$ (fine structure constant), and allow for solutions with huge values of $\mathcal{Q}^2$. So the only possiblity is to violate \eqref{dScond}, that is, to have
\begin{equation} \sqrt{\left(\frac{f'}{f}\right)^2+ \left(\frac{\theta'}{4\pi^2 f}\right)^2}\leq \left\vert\frac{V'}{V}\right\vert.\label{dScond2}\end{equation}
Thus, we cannot have a large dependence on the quintessence field on either the QED axion decay constant, or the gauge coupling itself. Since the quintessence field is rolling, this translates to a constraint on the time dependence of the fine structure constant as well as that of the would-be axion decay constant (we emphasize however that $\theta$ has not even been measured to be nonzero). 

Ignoring the $\theta$ dependence, we get the following result for the time variation of the axion decay constant, where we have introduced $c\equiv\vert V'/V\vert$ and used the slow-roll estimate  $\dot{\phi}\approx H$ as well as $f=1/\alpha$:
\begin{equation} \left\vert \frac{\dot{\alpha}}{\alpha}\right\vert\leq c\, H.\end{equation}
Using that $H\approx 10^{-10} (yrs)^{-1}$, we get\footnote{For $1/H = 14.4*10^{9} \,yrs$, one has $H=6.94 * 10^{-11}\, (yrs)^{-1}$.}
\begin{equation} \left\vert \frac{\dot{\alpha}}{\alpha}\right\vert\lesssim \frac{c}{10^{10}\, \text{years}},\end{equation}
and since experimentally we know that $c\lesssim1$ \cite{Ade:2015rim}, this is an upper bound for the time variation of $\alpha$.  Currently, the best experimental upper bound for the time variation of the fine structure constant comes from the Oklo natural nuclear reactor \cite{Damour:1996zw}, which constraints 
\be
\frac{\dot{\alpha}}{\alpha}<10^{-15} (yrs)^{-1} \label{oklo}\,.
\ee
Therefore, our constraint is satisfied experimentally. We would need $c\sim10^{-5}$ to provide new meaningful bounds on $\dot{\alpha}/\alpha$.

We can also put bounds on any axion that couples to the QED axion \cite{Tanabashi:2018oca}. This is a periodic scalar field $\chi$ with a standard kinetic term that couples to the electromagnetic $U(1)$ field via a coupling
\begin{equation}g_{\chi}\frac{\chi}{8 \pi^2} F\wedge F\end{equation}
in the Lagrangian. We will also assume $\chi$ has a periodic potential, of the form
\begin{equation} V(\chi)= qV_0\left[1-\cos\left(\frac{\chi}{f_\chi}\right)\right].\end{equation}
Here, $q$ is a dimensionless parameter, and $f_\chi$ is the axion decay constant.

We will consider a situation where there is a rolling quintessence field, such as the one we considered so far, and also an axion $\chi$ coupling to electromagnetism.  In principle, both $q$ and $f_\chi$ could depend on $\phi$ as well, but we will assume they do not for simplicity.  As we will see, avoiding $dS_2$ minima leads to bounds on the axion mass and axion decay constant. Our formalism, and in particular equations \eqref{eq1}, \eqref{eq2} and \eqref{eq3} work in the same way in situations with more than one scalar. We just need to include equations like \eqref{eq1} for all the scalar directions we take into consideration. So in a scenario with quintessence and an additional axion field the constraint \eqref{oklo} gets modified to involve the axion mass and coupling $g_{\chi}$ in a nontrivial way.  We just need to apply \eqref{dScond2} to the axionic direction too. Explicitly, we obtain\footnote{In obtaining this constraint we have used that $q\gg1$, worked out the location of the minimum for any value of $\chi,\phi$, and evaluated at $\chi=\pi/2f_\chi$. We have also checked stability of the minimum with respect to $\phi$ and $\chi$, and that the resulting electric and magnetic charges are real and positive.  On top of this, the constraint \eqref{condition0} that ensures that the radius is positive must be enforced as well.} a constraint in terms of $\alpha=1/(4\pi f)$
\begin{equation} p\equiv \left\vert \frac{2\pi}{\alpha} f_\chi g_\chi \right\vert \leq \frac{\pi}{2\alpha^3}.\label{dls}\end{equation}
When the axion $\chi$ is the QCD axion, the parameter $p$ is directly related to the electromagnetic and color anomalies  of the axial current associated with the axion, $E$ and $N$ \cite{Tanabashi:2018oca}\footnote{To compare to \cite{Tanabashi:2018oca}, one must use that the normalization in that reference is that $\int F_{\text{PDG}}=2\pi\sqrt{f}n$ (see Sec. 116, review on magnetic monopoles), where $n$ is an integer. So $F_{\text{PDG}}=\frac{1}{g}$, and our $\theta$ term becomes $ \frac{g_\chi\alpha \chi}{2\pi}\int F_p\wedge F_p$, so that $G_{A\gamma\gamma}=\frac{g_\chi\alpha}{\pi}$.}
\begin{equation} p= \frac{\pi}{\alpha}\left(\frac{E}{N}-1.92\right).\label{gc}\end{equation}
Equation \eqref{dls} then puts a bound on $E/N$, that it should be lower than $\sim 9400$. This is way above what any reasonable model attains \cite{Tanabashi:2018oca}, so it is not a very interesting constraint. 

Our constraints also apply to models of axion quintessence \cite{Carroll:1998zi}, but they are trivially satisfied, since the axion decay constant must be transplanckian. Of course, this kind of models is in trouble anyway with the axion version of the WGC (see e.g. the review \cite{Palti:2019pca}).

\section{Discussion}\label{Sec:discussion}
In this note we have pointed out that a simple flux compactification ansatz can lead to stable, parametrically controlled de Sitter solutions starting from a theory with a runaway scalar potential and abelian $p$-form gauge fields, provided the simple constraint \eqref{dshigh} involving the slope of the potential and gauge kinetic functions is satisfied. The $dS_n$ solutions we get are perturbatively unstable for $n>2$, but perturbatively stable for $n=2$ This corresponds to the interesting case of compactifications of a four-dimensional theory.

We then check in three simple stringy examples, not covered by usual  Maldacena-Nu{\~n}ez  arguments, how this constraint fails to be satisfied. Indeed, if any version of the no-dS conjecture actually holds in string theory, then the converse of \eqref{dshigh} must follow as well. We find that \eqref{dshigh} is satisfied even in the higher-dimensional cases where our construction would produce an unstable de Sitter solution anyway. Perhaps, as \cite{Andriot:2018wzk,Ooguri:2018wrx,Andriot:2018mav} suggests, even mildly tachyonic de Sitter is problematic string theory, at least close to perturbative corners. It could also be that the constraint is satisfied because we might be able to further compactify our solutions to $dS_2$ solutions. We leave this to future work.

Applying our constraint to a quadratic potential for a canonically normalized field (such as any stabilized modulus or KK mode in a string compactification) leads to interesting consequences. If the vacuum is Minkowski, then the dS minimum is at a Planckian distance, so one would not trust it anyway. If the vacuum is AdS, then one can have a de Sitter minimum at a subplanckian distance only if the lightest scalar has a mass that is parametric above the AdS scale (and the consistent truncation we work in is valid). This goes agains the AdS moduli separation and AdS-KK scale separation conjectures \cite{Gautason:2018gln,Lust:2019zwm}, and so we also do not know of stringy examples where this works. If one finds a parametric scale-separated AdS with a light $p-1$-form gauge field, or more generally any AdS solution where our truncation (with a value of the mass above the AdS scale) is valid, and our techniques automatically provide a (perturbatively unstable, except in two dimensions) dS solution.

In the second part of our note, we have explored the phenomenological consequences of the no dS conjecture coupled to our result. We have been able to constrain both the time-variation of the fine structure constant in a quintessence scenario, as well as any axion-photon coupling (including the QCD axion). The no dS conjecture implies $\dot{\alpha}/\alpha <10^{-10}\text{year}^{-1}c$, where $c$ is the slope of the quintessence potential, while current experimental bounds are at $\dot{\alpha}/\alpha <10^{-15}\text{year}^{-1}$. We consider this very interesting since it demonstrates the self-consistency of the no-dS conjecture; the problematic running of the fine-structure constant in generic quintessence models is strongly constrained by the conjecture.  We have also explored constraints for the coupling of any axion to electromagnetism. In particular, for the QCD axion, we obtain weak constraints that are easily accommodated by any reasonable GUT. 

The first part of our note is just a general way of constructing de Sitter solutions, and is independent of the validity of the no dS conjecture or any of its variants. But the phenomenological implications really depend on the validity of the statement \cite{VanRiet:2011yc, Danielsson:2018ztv, Brennan:2017rbf, Obied:2018sgi, Ooguri:2018wrx}. If this conjecture is wrong so are all conclusions drawn from this, including the simple observation made in this note.\footnote{Strictly speaking we demanded something more than the specific ``no-dS'' conjecture as stated in \cite{Danielsson:2018ztv, Brennan:2017rbf, Obied:2018sgi, Ooguri:2018wrx}, we extend it to the non-existence of meta-stable dS vacua in $D<4$. Even if higher-dimensional de Sitter is somehow pathological, it may be that lower-dimensional solutions are OK. Indeed, \cite{Maldacena:2019cbz} studies a version of the SYK model which is holographically dual to JT gravity in a $dS_2$ spacetime (although it is unclear whether the theory is Einstein). However if we consider the lack of trustworthy dS constructions from string compactifications as a reason for the conjecture as in \cite{VanRiet:2011yc, Danielsson:2018ztv}, then the no-dS conjecture in two dimensions is as plausible as the four-dimensional case.  }

We have also discussed the subtleties of the classical stability analysis of a $dS_2$ solution. Although perturbations in the homogeneous mode are stable, there are inhomogeneous perturbations that are classically unstable, leading to ``Bousso beading''. These inhomogeneities clump and collapse, leading to a crunch, while away from the inhomogeneities the solution is still $dS_2$. We interpret these as pair production of 2d black holes by the expanding $dS_2$ background, and thus not an actual vacuum instability, although of course we cannot be sure this is correct.  Finally, we also have to address quantum mechanical instabilities. Since the dS$_2\times S^2$ solutions can be seen as the boundary of the space of black hole solutions in 4D it should not come as a surprise that these solutions are quantum mechanically unstable with respect to Hawking radiation and Schwinger pair production. Unlike the classical instability described before, this can happen even in the homogeneous sector. An in-depth study of this was recently carried out by us in \cite{Montero:2019ekk}, where it was found that the lifetime of the two-dimensional de Sitter solution is always finite, and we argued that avoidance of a certain crunch singularity leads to the $dS_2$ being metastable (in the sense that it gets a lifetime that grows exponentially as the gauge coupling of the gauge theory goes to zero). 

In the allowed regime, these quantum instabilities are essentially bubble nucleation instabilities of 4D meta-stable dS vacua with the role of the Brown--Teitelboim bubbles replaced by electron-positron pairs. So our instabilities in 2d are still such that we can think of the $dS_2$ solutions (away from the clumps of the beading process) as metastable.

The take-home message of this note is that one can often get controlled lower-dimensional de Sitter solutions from runaway potentials. One may regard this as a simple setup where one can explicitly construct de Sitter solutions in string theory, or as a way to place certain runaway potentials in the Swampland. It will be important to figure out which attitude is correct, if either.

\appendix

\section*{Acknowledgements}
We thank Irene Valenzuela,  Bartosz Fornal, Prateek Agrawal, Gary Shiu and especially Luis Iba\~{n}ez for useful discussions and comments. The work of TVR and GV is supported by the FWO Odysseus grant G.0.E52.14 and the C16/16/005 grant of the KULeuven. MM is supported by Grant 602883 from the Simons Foundation.

\section{Equations of motion}\label{App:EOM}
Here we derive the equations of motion \eqref{eq1}-\eqref{eq3} in the main text. Using the ansatz \eqref{Ansatz}, and only allowing time dependence of the fields, the action becomes (after integration by parts)

\begin{equation}S=\int dt\left(2 h - R^2\frac{d}{dt}{\left(\frac{\dot{h}}{h}\right)} -2\dot{R}^2+\frac{h R^2}{2}\dot{\phi}^2-\frac{fR^2}{2}\left(\frac{E^2}{h}-B^2 h\right)-V h R^2+\frac{\theta}{4\pi^2} EB R^2 \right),\label{act}\end{equation}
where $E$ is the electric field, and $B$ is the magnetic field. It has the value $E=\frac{Q_e h}{R^2f}$. The magnetic field is quantized, $B=Q_m/R^2$. These two conditions are enforced off-shell in two dimensions since the gauge field has no dynamical degrees of freedom. The resulting action is
\begin{equation}S=\int dt\left(2 h -R^2\frac{d}{dt}{\left(\frac{\dot{h}}{h}\right)} -2\dot{R}^2+\frac{ h R^2}{2}\dot{\phi}^2-\frac{h}{2R^2}\mathcal{Q}^2-V h R^2 \right).\label{act2}\end{equation}

 Setting $\dot{R}=0$ and $\dot{\phi}=0$ since we are looking for static solutions, the equation of motion for $R^2$ is just
\begin{equation}\mathcal{R}_2=\frac{1}{h}\frac{d}{dt}{\left(\frac{\dot{h}}{h}\right)}=V-\frac{\mathcal{Q}^2}{2R^4}.\label{dsg}\end{equation}
In the above, the left hand side is precisely the curvature of the two-dimensional manifold, $\mathcal{R}_2$, so we recover the main text.  The equation of motion for $h$, which one obtains by first integrating by parts the second term in \eqref{act2}, is
\begin{equation}\frac{2}{R^2}-\frac{\mathcal{Q}^2}{2R^4}-V=0,\end{equation}
or, multiplying by $-R^4$, 
\begin{equation}VR^4-2R^2+\frac{\mathcal{Q}^2}{2}=0,\label{rrrr}\end{equation}
which is the equation in the main text. Finally, the equation of motion for $\phi$ is
\begin{equation}0=V'+\frac{(\mathcal{Q}^2)'}{2R^4}.\label{eqphi2}\end{equation}
In case we deal with a truncation to several scalars, as we do in the axion case in the main text, \eqref{eqphi2} becomes a vector equation where the prime denotes derivative with respect to every scalar. We do not expect to find a solution in situations with more than two scalars.

As for the stability analysis, we can just refer to Appendix C.3 of \cite{Hamada:2017yji}. We just need to ensure that the second derivative of the scalar potential with respect to the quintessence field is positive. One obtains for a black hole with $Q_m = m Q_{tot}$ and $Q_e = (1-m) Q_{tot}$
\begin{equation}V''>V' \left( \frac{f''}{f} - 2 \frac{(1-m)^2 f'}{(1-m)^2 f + m^2 f^3} \right)\,.\end{equation}
For a magnetic black hole $m=1$ and exponential dependence $f=e^{\delta\phi}$, $V=V_0e^{-\gamma\phi}$ this becomes
\begin{equation}\gamma^2 > - \gamma \delta,\end{equation}
and so is satisfied for any positive values of $\gamma$ and $\delta$.

For an electric black hole $m=0$ and exponential dependence $f=e^{\delta\phi}$, $V=V_0e^{\gamma\phi}$ (note the change in the definition of $V$) one instead obtains
\begin{equation}
\gamma^2 > - \gamma \delta\,,
\end{equation}
which is again satisfied for positive $\gamma$ and $\delta$. Electric and magnetic black holes are thus both stable in the regime where they have solutions.

\section{Higher dimensions}\label{app:higherdim}
The main idea in this note is that, when compactifying to two dimensions a higher-dimensional solution with a runaway potential and a nontrivial gauge kinetic function satisfying certain constraints, de Sitter solutions are unavoidable. Here we generalize the picture to higher dimensions. One can consider a general $D$-dimensional solution with an abelian $p$-form fieldstrength and a scalar with an action of the form
\begin{equation}S=\frac{1}{2\kappa_{D}^2}\int d^{D}x\, \sqrt{\vert g_E^{10}\vert}\left[ R_E-\frac12(\partial\phi)^2-f(\phi)\vert F_p \vert^2-V(\phi)\right],\end{equation}
where without loss of generality we can take $\gamma>0$ and $\delta>0$ (if $\delta$ is negative for a certain $F_p$, it will be positive for its electromagnetic dual). 

Following \cite{Bremer:1998zp}, we can now reduce on a product manifold $X\times Y$, of dimensions $d_x,d_y$, $d_x+d_y=D$ and $d_X>2$, and with metric
\begin{equation} ds_{D}^2= e^{2\alpha\rho}ds_{X}^2+ e^{2\beta\rho}ds_{Y}^2,\end{equation}
where
\begin{equation}\alpha^2\equiv\frac{d_y}{2(d_x-2)(d_x+d_y-2)},\quad \beta=-\frac{d_x-2}{d_y}\alpha.\end{equation}
This ansatz ensures that the scalar $\rho$ has a canonical kinetic term after dimensional reduction, and that the lower-dimensional theory is in the Einstein frame. The kinetic term of both scalars is
\begin{equation} \mathcal{L}_{\text{scalars}}^{\text{kin.}}=\frac12\left[ (\partial \rho)^2+e^{2\alpha\rho}(\partial\phi)^2\right].\label{lkin}\end{equation}

We will consider the particular case where $Y=S^p$ is a sphere threaded by $p-$form flux, and only look at a magnetic solution; the electric one can be obtained by electric-magnetic duality. The resulting $d_x$-dimensional effective action has two scalars $\phi$ and $\rho$, and a potential
\begin{equation} V_{X}=-d_y(d_y-1)e^{2(\alpha-\beta)\rho}+\frac{Q^2}{2}e^{2(d_x-1)\alpha\rho}f(\phi)+V(\phi)e^{2\alpha\rho}.\end{equation}
There is only one critical point, where $V$ takes the value
\begin{equation}V_{X}^{\text{min.}}=\frac{(d_x-2) e^{\frac{d_y \rho }{(d_x-2) (d_x+d_y-2)}} \left((d_y-1) f  V' +V  f' \right)}{(d_x+d_y-2) f' }.\end{equation}
Demanding that this is positive leads to \eqref{dshigh} in the main text. The charge is determined by 
\begin{equation}Q^2=-\frac{2 e^{-2 \alpha d_x-1 \rho } V' }{f' }.\end{equation}
In this case there is no obstruction akin to \eqref{condition0}; in other words, there is no bound on the charge the solutions can attain. In the $d_x=2$ case, the upper bound \eqref{condition0} can be interpreted as the condition for an ``ultracold black hole'' \cite{Romans:1991nq}, and it is related to the maximum charge that can one can fit on a static patch. For $d_x>2$ there is no such interpretation since the would-be brane whose near horizon geometry we are computing has to stretch out of the static patch anyway. 

Just as in the two-dimensional case, we need to study stability of the solution. As before, we will compute one diagonal element and the determinant of the Hessian $\mathcal{H}$.  We work in the specific case of $V=V_0e^{-\gamma\phi}$ and $f=e^{\delta\phi}$. The stability of the solution is then determined by
\begin{align} \partial^2_\phi V&=\frac{1}{2} \delta  Q^2 (\gamma +\delta ) e^{\delta  \phi +(d_x-1) \rho },\nonumber\\ \text{det}(\mathcal{H})&=-\frac{\delta  (d_x-2) Q^4 (\delta +\gamma  (d_x-1)) (\gamma(1-d_y) +\delta) e^{2 \delta  \phi +2 (d_x-1) \rho }}{4 \gamma  d_y}.\end{align}
Taking into account \eqref{dshigh}, we see that the determinant is always negative. Thus, the solution is always unstable. Presumably, we also get similar instabilities for more general $f$ and $V$. Note that, as discussed earlier, this conclusion does not apply to the $dS_2$ case, as here the sphere radius is fixed by the Hamiltonian condition.

We can compute explicitly the mass of the tachyonic direction, to then compare to the refined no dS conjecture \cite{Andriot:2018wzk,Ooguri:2018wrx,Andriot:2018mav}. The general expression is not illuminating, so we will not write it down here, but at the threshold \eqref{dshigh} it is 
\begin{equation} \frac{\vert m_{\mathcal{T}}^2\vert}{V_{X}^{\text{min.}}}=\frac{\delta^2 (d_x+d_y-2)^2}{(d_x-2)^2 (d_y-1)^2+\delta ^2 d_y^2}.\end{equation}
This is an $\mathcal{O}(1)$ quantity, for $\delta$ of order 1. So whether or not this satisfies \cite{Ooguri:2018wrx} depends on the precise value of the parameter $c'$ that one takes.

\bibliographystyle{utphys}

\bibliography{refsTVR}

\end{document}